\newif\ifonecolumn
\onecolumnfalse
\ifonecolumn
  \documentclass[journal,draftclsnofoot,onecolumn, 12pt]{IEEEtran}
\else
  \documentclass[journal]{IEEEtran}
\fi

\usepackage[T1]{fontenc}

\usepackage{amsmath}
\usepackage{cite}
\usepackage{array}
\usepackage{stfloats}
\usepackage{url}
\usepackage{amsmath,amssymb}
\usepackage{epsfig,fancyhdr}
\usepackage{CJKutf8}
\usepackage{times}
\usepackage{graphicx}
\usepackage{caption}
\usepackage{subcaption}
\usepackage{multirow}
\usepackage{stfloats}
\usepackage{color}
\usepackage[dvipsnames]{xcolor}
\usepackage{threeparttable}

\usepackage{arydshln} 

\usepackage{bm}

\newcommand{\bms}[1]{\mbox{\footnotesize\boldmath$#1$}}

\newtheorem{thm}{Theorem}

\newtheorem{defin}{Definition}
\newtheorem{lemma}{Lemma}
\newtheorem{crly}[thm]{Corollary}
\newtheorem{eg}{Example}
\newtheorem{rmk}{Remark}

\begin{document}

\title{Sparse Zero Correlation Zone Arrays for Training Design in Spatial Modulation Systems}

\author{
Cheng-Yu~Pai,~\IEEEmembership{Member,~IEEE}, 
Zilong Liu,~\IEEEmembership{Senior Member,~IEEE},
and Chao-Yu~Chen,~\IEEEmembership{Senior Member,~IEEE}
%
  }

\maketitle

\begin{abstract}
This paper presents a novel training matrix design for spatial modulation (SM) systems, by introducing a new class of two-dimensional (2D) arrays called sparse zero correlation zone (SZCZ) arrays. An SZCZ array is characterized by a majority of zero entries and exhibits the zero periodic auto- and cross-correlation zone properties across any two rows. With these unique properties, we show that SZCZ arrays can be effectively used as training matrices for SM systems. Additionally, direct constructions of SZCZ arrays with large ZCZ widths and controllable sparsity levels based on 2D restricted generalized Boolean functions (RGBFs) are proposed. Compared with existing training schemes, the proposed SZCZ-based training matrices have larger ZCZ widths, thereby offering greater tolerance for delay spread in multipath channels. Simulation results demonstrate that the proposed SZCZ-based training design exhibits superior channel estimation performance over frequency-selective fading channels compared to existing alternatives.   
\end{abstract}


\section{Introduction}
Zero correlation zone (ZCZ) sequence sets were initially introduced in \cite{Fan99}. These sequences are characterized by their zero periodic auto- and cross-correlation functions for certain time-shifts around the in-phase position. Throughout the years, numerous research attempts have been made on ZCZ sequence design \cite{Tang10,Torii04, Zhou08,Liu14,Super_182,Pai_23}. From the application standpoint, ZCZ sequences can be employed  as training sequences for optimal channel estimation performances in multi-input and multi-output (MIMO) systems \cite{Yuan_05,Yang_02,Hu_17}, spreading sequences in the quasi-synchronous code-division multiple-access (QS-CDMA) system \cite{Long98}, as well as pilot sequences for frequency synchronization and interference avoidance in multi-cell systems \cite{Zhang_10,Zhang_12}.

On the other hand, in recent years, spatial modulation (SM) has emerged as a promising multiple-antenna technique due to its unique feature that only one transmit antenna (TA) is activated per time-slot \cite{SM_2,SM_3,SM_4,SM_6,SM_5}. Thanks to this feature, SM enjoys a number of advantages, such as zero inter-antenna interference over flat-fading channels, reduced complexity, and enhanced energy efficiency. Since only one TA is activated per time slot,  channel estimation in SM systems requires sparse training matrices with each column containing only one non-zero entry. This contrasts with the traditional dense matrices used in conventional MIMO systems. A number of studies on channel estimation for SM systems over flat-fading channels have been conducted \cite{SM_Flat_2,SM_Flat_3,SM_timevaring_4,SM_timevaring_5}. For SM channel estimation over frequency-selective channels, Liu {\it et al.} introduced the cross Z-complementary pairs (CZCPs) \cite{CZCP-1st} each having 1) front-end and tail-end ZCZs for their aperiodic autocorrelation sums and 2) tail-end ZCZ for their aperiodic cross-correlation sums. Due to these sequence pair properties, it is demonstrated that inter-symbol interference (ISI) and inter-antenna interference (IAI) can be mitigated when the delay spread is smaller than the relevant ZCZ widths of the CZCPs. Since then, various constructions of CZCPs have been developed \cite{Fan_20, Adhikary_20, Huang_20, Yang_21, Zeng_22, Das_22_ISIT, Zhang_22}. As a further generalization, the CZCP concept has also been extended to the cross Z-complementary set (CZCS) \cite{Huang_22,Huang_22_ISIT}. However, the existing state-of-the-art primarily follows the training framework proposed by \cite{CZCP-1st}, where several kernel CZCPs (or CZCSs) are arranged in a specific pattern to generate SM-compatible regular sparse matrices. Consequently, the ZCZ widths of these training matrices are limited by the kernel CZCPs or CZCSs. 

Motivated by the aforementioned background, this paper presents a novel training framework for SM systems by introducing a new class of two-dimensional (2D) arrays called the sparse ZCZ (SZCZ) array. Our goal is to go beyond the existing kernel based approaches so as to offer a greater flexibility in SM training design. Formally, each SZCZ array is dominated by zero elements, and any two rows of such an SZCZ array exhibit the zero periodic auto- and cross-correlation zone property.
By fully exploiting the sparsity and correlation properties of SZCZ arrays, we show that these SZCZ arrays can be directly used as training matrices for SM systems, thus referred to as SZCZ training matrices. It is worth mentioning that the ternary ZCZ arrays \cite{Chang_67,Xu_03,Wu_06} may not be feasible for the SM training matrices as they cannot guarantee the requirement of having only one non-zero entry per column. In this paper, we introduce the concept of 2D restricted generalized Boolean functions (2D RGBFs).
By carefully restricting certain variables in 2D RGBFs, we demonstrate that the positions of non-zero entries in each column of the corresponding 2D arrays can be controlled, which is crucial since only one non-zero entry is allowed for every SM training matrix column. Following this idea, we present direct constructions of SZCZ training matrices characterized by large ZCZ widths and controllable sparsity levels, utilizing 2D RGBFs. Compared with existing CZCP-based and CZCS-based training matrices, the proposed SZCZ training matrices possess ZCZ widths that are twice as large, providing greater tolerance for delay spread over frequency-selective channels. The simulation results demonstrate that the proposed SZCZ-based training design outperforms the performance of existing alternatives in channel estimation over frequency-selective fading channels.

The subsequent sections of this paper are organized as follows. In Section \ref{sec:background}, we introduce some notations and preliminaries, including the SM systems and the definitions of the SZCZ array. In Section \ref{sec:Training}, we propose a novel training framework for SM systems based on the SZCZ array and present the constructions of SZCZ arrays. Section \ref{sec:Simulation} provides the simulation results. Lastly, we provide the conclusion in Section \ref{sec:conclusion}.

\section{Preliminaries and Definitions}\label{sec:background}
 Throughout this paper, the following notations will be used:
\begin{itemize}
\item $\mathbb{Z}_q=\{0,1,\ldots,q-1\}$ represents the set of integers modulo a positive integer $q$;
\item $\xi=e^{2\pi \sqrt{-1}/q}$ is a primitive $q$-th root of unity;
\item we consider even integer $q$ in this paper;
\item $(\cdot)^{*}$ denotes the complex conjugation;
\item $(\cdot)_{b}$ represents the modulo $b$ operation where $b$ is the a positive integer;
\item $(\cdot)^T$ denotes the transpose;
\item $(\cdot)^H$ denotes the Hermitian of the matrix;
\item $\text{Tr}(\cdot)$ represents the trace of the matrix;
\item ``$+$'' and ``$-$'' denote $1$ and $-1$, respectively;
\item ${\bm 0}_{L}$ stands for an all-zero vector of length $L$.
\end{itemize}

Let $\{{\bm C}_0,{\bm C}_1,\ldots,{\bm C}_{N-1}\}$ denote a set of $N$ complex-valued sequences of length $L$ where
${\bm C}_g=(C_{g,0},C_{g,1},\ldots,C_{g,L-1})$ and $|C_{g,i}|\in \{0,1\}$ for $0 \leq g \leq N-1$.
The periodic cross-correlation function (PCCF) of sequences ${\bm C}_{g}$ and ${\bm C}_{k}$ at the time-shift $u$ is defined as
\begin{equation}
\theta({\bm C}_{g},{\bm C}_{k};u)=
\sum_{i=0}^{L-1}C_{g,(i+u)_L}C_{k,i}^{*},~u\neq 0.
\end{equation}
It is noted that $\theta({\bm C}_{g},{\bm C}_{k};u)=\theta({\bm C}_{k},{\bm C}_{g};-u)^{*}$. When $g=k$, it is referred to as the periodic autocorrelation function (PACF) of ${\bm C}_g$ denoted by $\theta({\bm C}_{g};u)$.
Additionally, the aperiodic cross-correlation function of sequences ${\bm C}_{g}$ and ${\bm C}_{k}$ at time-shift $u$ is defined as
\begin{equation}
  \rho({\bm C}_{g},{\bm C}_{k};u)=\begin{cases}
\sum\limits_{i=0}^{L-1-u}C_{g,i+u}C_{k,i}^{*}
&  0 \leq u\leq L-1;\\
\sum\limits_{i=0}^{L-1+u}C_{g,i}C_{k,i-u}^{*}, &  -L+1 \leq u<0.
\end{cases}
  \label{eq:ap_cross}
\end{equation}
Likewise, the aperiodic autocorrelation function of ${\bm C}_g$ is denoted as $\rho({\bm C}_g;u)$.
\begin{defin}
\cite{Huang_22} For positive integers $N$ and $Z$ with $Z\leq L$, denote two intervals as
\[
\mathcal{R}_1\triangleq\{1,2,\ldots,Z\}~\text{and}~\mathcal{R}_2\triangleq\{L-Z,L-Z+1,\ldots,L-1\}.
\]
Then, a set of $N$ polyphase sequences $\{\bm C_0, \bm C_1, \ldots, \bm C_{N-1}\}$ of length $L$ is called an $(N,L,Z)$-CZCS if it satisﬁes
the following conditions.
\begin{equation}
\begin{cases}
\sum \limits _{i=0}^{N-1}{\rho} ({\bm C}_i; u) =0,&\text{for all}\,\,|u |\in \left(\mathcal {R}_1 \cup \mathcal {R}_2\right)\cap \mathcal {R}; \\
\sum \limits _{i=0}^{N-1}{\rho} ({\bm C}_i,{\bm C}_{(i+1)_{N}}; u) =0,& \text{for all}\,\,|u | \in \mathcal {R}_2
\label{eq:C1&C2}
\end{cases}
\end{equation}
where $\mathcal {R}\triangleq\{1,2,\ldots, N-1\}$.

By setting $N=2$, a CZCS is reduced to a CZCP \cite{CZCP-1st}, referred to as a $(L,Z)$-CZCP where $Z\leq L/2$ when $L$ is even. An $(L,L/2)$-CZCP is also known as a perfect CZCP.
\end{defin}

Next, we formally introduce the concept of an SZCZ array.
\begin{defin}
Denote the 2D array $\bm C$ of size $N\times L$ as
\begin{equation}
\bm C=
\begin{pmatrix}
{\bm C}_0\\
{\bm C}_1\\
\vdots \\
{\bm C}_{N-1}
\end{pmatrix}=
\begin{pmatrix}
C_{0,0}& C_{0,1} & \cdots & C_{0,L-1}\\
C_{1,0} & C_{1,1} & \cdots & C_{1,L-1}\\
\vdots & \vdots & \ddots & \vdots   \\
C_{N-1,0} & C_{N-1,1} & \cdots & C_{N-1,L-1}
\end{pmatrix},
\end{equation}
where each row sequence ${\bm C}_g$ has $M$ non-zero entries.
The array $\bm C$ is said to be an $(N,L,Z,\mathcal{S})$-SZCZ array if the following conditions hold:
   \begin{equation}
      \begin{aligned}
        &\theta(\bm{C}_{g},\bm{C}_{k};u)\\&=
        \begin{cases}
          M,  & u=0, 0 \leq g=k \leq N-1; \\
          0, & 1\leq|u|\leq Z,~0 \leq g=k \leq N-1; \\
          0, & |u|\leq Z,~ 0 \leq g\neq k \leq N-1,
        \end{cases}
        \label{eq:SZCZ_seq}
      \end{aligned}
    \end{equation}
where $Z$ represents the width of the ZCZ and $\mathcal{S}=(L-M)/L$ indicates the sparsity level. If $\mathcal{S}=0$, the array $\bm C$ is reduced to the traditional dense ZCZ array. 
\end{defin}

In Section \ref{sec:Training}, we will show that the SZCZ arrays play a crucial role in SM training design. 
\subsection{2D Restricted Generalized Boolean Functions}\label{sec:RGBF}
In this subsection, we introduce the concept of the 2D RGBFs, which serves as a useful tool for constructing the SZCZ array. To this end, let us first introduce the 2D GBF \cite{Pai_22}. A 2D GBF $f$: $\mathbb{Z}^{n+m}_2$ $\rightarrow$ $\mathbb{Z}_q$ comprise $n+m$ variables $y_1,y_2,\ldots,y_n,x_1,x_2,\ldots,x_m$ where $y_g,x_i \in \{0,1\}$ for $1\leq g\leq n$ and $1\leq m\leq m$. The 2D array ${\bm f}$ associated to the 2D GBF $f$ is represented by 
\begin{equation}
{\bm f}=
\begin{pmatrix}
f_{0,0}& f_{0,1} & \cdots & f_{0,2^{m}-1}\\
f_{1,0} & f_{1,1} & \cdots & f_{1,2^{m}-1}\\
\vdots & \vdots & \ddots & \vdots   \\
f_{2^{n}-1,0} & f_{2^{n}-1,1} & \cdots & f_{2^{n}-1,2^{m}-1}
\end{pmatrix},
\end{equation}
where $f_{g,i}=f((g_1,g_2,\cdots,g_n),(i_1,i_2,\cdots,i_m))$, $g=\sum_{h=1}^{n}g_{h}2^{h-1}$, and $i=\sum_{j=1}^{m}i_{j}2^{j-1}$.
The corresponding complex-valued array is provided by
$\xi^{\bms f}=(\xi^{f_{g,i}})$ for $g=0,1,\ldots,2^{n}-1$ and $i=0,1,\ldots,2^{m}-1$.
\begin{eg}\label{eg:2Dgbf}
Considering $q=2,n=2$, and $m=3$, let $f=x_{1}x_2+y_{1}x_3+y_2$. The associated complex-valued array $\xi^{\bm f}$ is given by
\begin{equation*}\label{eg:2DGBF_F}
\xi^{\bm f}=
\begin{pmatrix}
+& + & + & - & +& + & + & -\\
+& + & + & - & -& - & - & +\\
-& - & - & + & -& - & - & +\\
-& - & - & + & +& + & + & -
\end{pmatrix}.\end{equation*}
\end{eg}

Denote a set of $p$ indices by $W=\{w_1,w_2,\ldots,w_p\}\subset \{1,2,\ldots,n\}$. Let $\bm x=(x_{w_1},x_{w_2},\ldots,x_{w_p})$ and ${\bm y}=(y_{1},y_{2},\ldots,y_{p})$ where $y_{l}\in \{0,1\}$. A 2D RGBF $f|_{\bms x={\bms y}}$ is defined by restricting variables ${\bm x}$ in the $f$ to the certain known ${\bm y}$. 
To simplify, consider the associated array ${\bm f}|_{{\bms x}={\bms y}}$ as the complex-valued array with component equal to $\xi^{f_{g,i}}$ if  $i_{w_{\alpha}}=g_{\alpha}$ for $\alpha=1,2,\ldots,p$ and equal to zero otherwise. Therefore, the associated 2D array ${\bm f}|_{{\bms x}={\bms y}}$ is a sparse array of size $2^{n} \times 2^{m}$.
\begin{rmk}\label{rmk:RGBF_sparsity}
In a sparse array ${\bm f}|_{{\bms x}={\bms y}}$ of size $2^{n} \times 2^{m}$, the number of non-zero entries is $2^{m+n-p}$. Thus, the sparsity is given by 
\begin{equation}
\mathcal{S}=\frac{2^{m+n}-2^{m+n-p}}{2^{m+n}}=\frac{2^{p}-1}{2^p}.
\end{equation}
\end{rmk}
\begin{eg}
Using the same notations as provided in {Example \ref{eg:2Dgbf}}, let $p=2$ and $\{w_1,w_2\}=\{1,2\}$. Then we set ${\bm x}=(x_1,x_2)$ and ${\bm y}=(y_1,y_2)$. The associated array ${\bm f}|_{{\bms x}={\bms y}}$ is given by 
\begin{equation*}
{\bm f}|_{\bms x={\bms y}}=
\begin{pmatrix}
+& 0 & 0 & 0 & +& 0 & 0 & 0\\
0& + & 0 & 0 & 0& - & 0 & 0\\
0& 0 & - & 0 & 0& 0 & - & 0\\
0& 0 & 0 & + & 0& 0 & 0 & -
\end{pmatrix}\end{equation*}
with the sparsity $\mathcal{S}=3/4$.
\end{eg}
\subsection{Spatial Modulation}
\begin{figure*}[htbp]
	\centering
	\includegraphics[width=140mm]{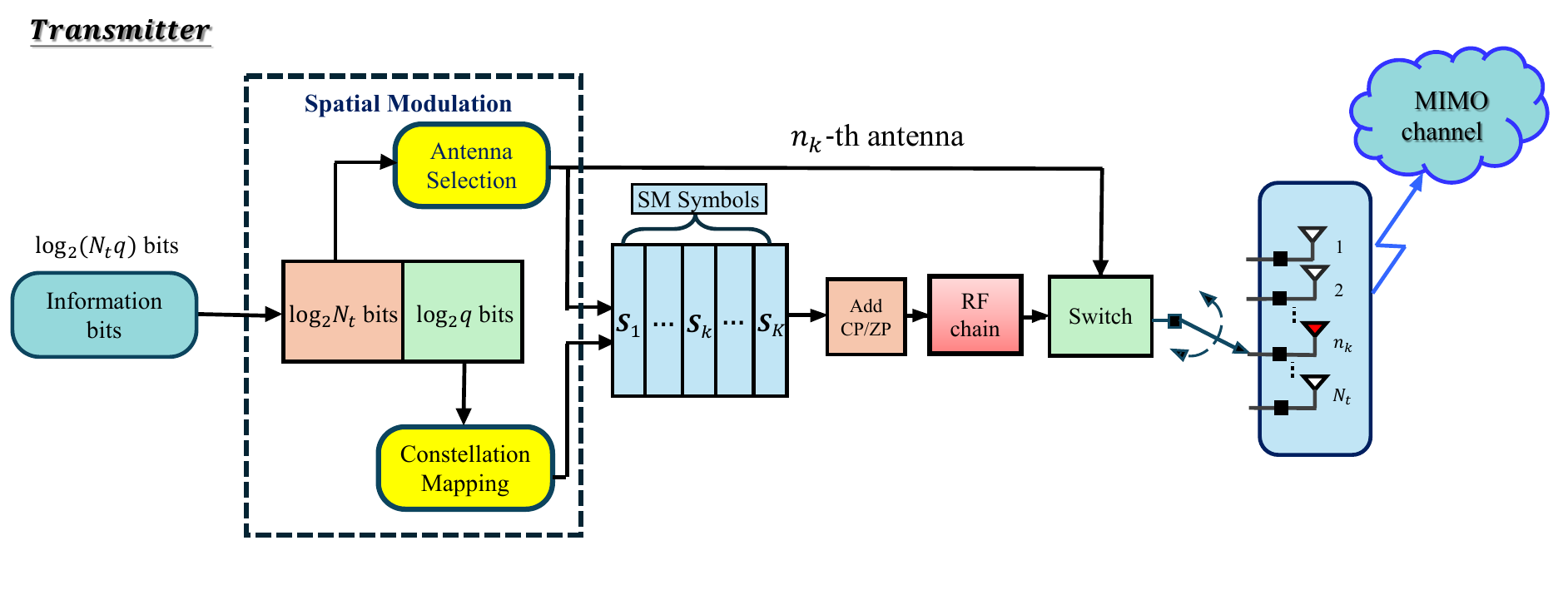}
	\caption{Generic transmitter structure of SC-SM systems.}
    \label{fig:transmitter}	
\end{figure*}

Consider a broadband single-carrier spatial modulation (SC-SM) system with $N_t$ transmit antennas (TAs), $N_r$ receive antennas (RAs), and a QAM/PSK modulation scheme, where the constellation size is denoted by $q$, operating over frequency-selective fading channels, as shown in Fig. \ref{fig:transmitter}. The SM symbol ${\bm S}_k$, for $0\leq k\leq K$, comprises $\log_2(N_t q)$ information bits. Specifically, $\log_2{N_t}$ bits are used for activating one of $N_t$ TAs and $\log_{2}q$ bits are used to select a QAM/PSK symbol $S_{k}$ transmitted by the activated TA at time slot $k$. Let $n_k$ denote the index of the selected TA at time slot $k$ and hence the SM symbol ${\bm S}_k$ can be expressed as 
\begin{equation}
{\bm S}_k=[\underbrace{0, \ldots, 0}_{n_k - 1}, S_{k}, \underbrace{0, \ldots, 0}_{N_t - n_k}]^T.
\end{equation}
Further details on a broadband SC-SM system can be found in \cite{SM_4,SM_6}.
\section{Training Framework for SM Systems}\label{sec:Training}
In this section, we begin by introducing the system model and outlining the design criteria for the training matrix in SC-SM systems. Subsequently, we propose a novel training matrix design utilizing SZCZ arrays.
\subsection{Training Matrix Design}\label{sec:system_model}
\begin{figure}[htbp]
	\centering
	\includegraphics[width = 85mm]{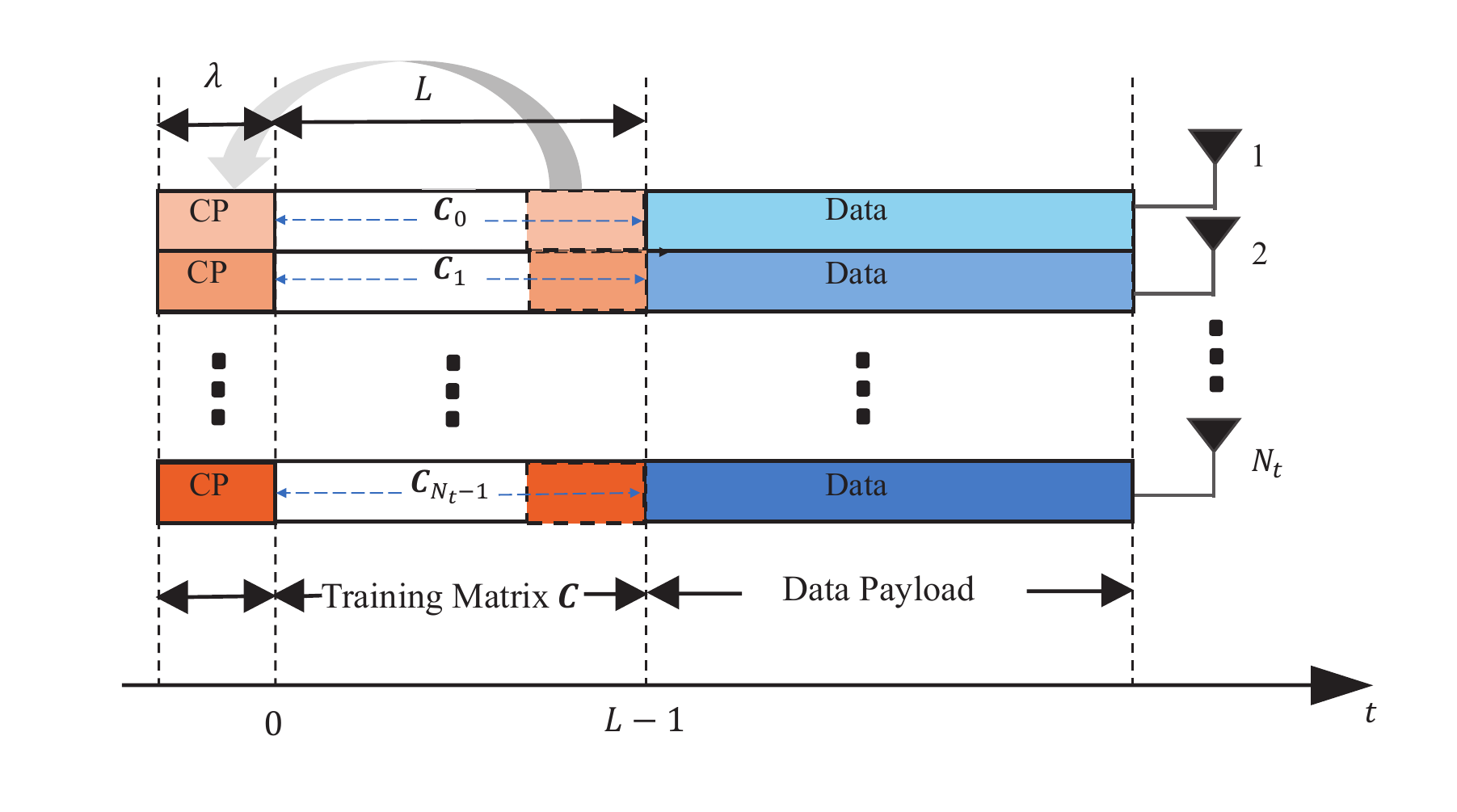}
	\caption{A training-based multiple-antenna transmission structure.}
    \label{fig:training}	
\end{figure}

For a generic training-based multiple-antenna transmission structure shown in Fig. \ref{fig:training}, let us consider $N_t$ TAs and $N_r$ RAs. Before transmitting the data payload, the training sequences ${\bm C}_0$, ${\bm C}_1,\ldots,$ ${\bm C}_{N_t-1}$ are sent  from $N_t$ TAs for estimating the channel state information (CSI) at the receiver. A cyclic prefix (CP) is added before each training sequence for ISI mitigation in multipath channels. These training sequences can be expressed as the matrix form, referred to as the training matrix.
Specifically, the training matrix is given by
\begin{equation}\label{eq:training}
\bm C=
\begin{pmatrix}
{\bm C}_0\\
{\bm C}_1\\
\vdots \\
{\bm C}_{N_t-1}
\end{pmatrix}=
\arraycolsep=0.8pt
\begin{pmatrix}
C_{0,0}& C_{0,1} & \cdots & C_{0,L-1}\\
C_{1,0} & C_{1,1} & \cdots & C_{1,L-1}\\
\vdots & \vdots & \ddots & \vdots   \\
C_{N_t-1,0} & C_{N_t-1,1} & \cdots & C_{N_t-1,L-1}
\end{pmatrix},
\end{equation}
where ${\bm C}_g=(C_{g,0}, C_{g,1},\ldots,C_{g,L-1})$ is transmitted over the $g$-th antenna with equal energy $\sum_{l=0}^{L-1}|C_{g,l}|^2=M~~\text{for}~0\leq g\leq N-1$. Since only one TA is activated per time slot in SM system, each column of the training matrix contains only one non-zero entry. This implies that each column $(C_{0,i},C_{1,i},\ldots,C_{N-1,i})^T$ of (\ref{eq:training}) has one non-zero entry for $0\leq i\leq L-1$.   
The channel matrix with $(\lambda+1)$-multipath is denoted as
\begin{equation}\label{eq:channel}
{\bm H}=
({\bm H}_0,{\bm H}_1,\ldots,{\bm H}_{\lambda})_{N_r\times N_t(\lambda+1)}
\end{equation}
where ${\bm H}_l\in \mathbb{C}^{N_r\times N_t}$ denotes the $l$-th multipath channel submatrix with its constituent entries drawn from $\mathcal{C}\mathcal{N}(0,\,1/(\lambda+1))$.
Besides, let
\begin{equation}\label{eq:circular_matrix}
{\bm X}=
\begin{pmatrix}
{\bm C}^{(0)}\\
{\bm C}^{(1)}\\
\vdots \\
{\bm C}^{(\lambda)}
\end{pmatrix}_{N_t(\lambda+1)\times L}
\end{equation}
where ${\bm C}^{(u)}$ represents the $u$-th right cyclic shift of the matrix ${\bm C}$ in (\ref{eq:training}), i.e., 
\begin{equation}
{\bm C}^{(u)}=
\begin{pmatrix}
C_{0,L-u}& C_{0,L-u+1} & \cdots & C_{0,L-u-1}\\
C_{1,L-u} & C_{1,L-u+1} & \cdots & C_{1,L-u-1}\\
\vdots & \vdots & \ddots & \vdots   \\
C_{N_t-1,L-u} & C_{N_t-1,L-u+1} & \cdots & C_{N_t-1,L-u-1}
\end{pmatrix}.
\end{equation}
The received signal ${\bm Y}\in \mathbb{C}^{N_r\times L}$ is 
\begin{equation}
\bm Y={\bm H}{\bm X}+{\bm V}
\end{equation}
where ${\bm V}\in \mathbb{C}^{N_r\times L}$ is the complex additive white Gaussian noise (AWGN) with each entry having a distribution from $\mathcal{C}\mathcal{N}(0,\,\sigma^{2}_v)$. The signal-to-noise ratio (SNR) at each RA is given as $1/\sigma^{2}_v$. By utilizing the least-square (LS) channel estimator \cite{Yang_02}, the estimated channel matrix is given by
\begin{equation}
\hat{\bm H}={\bm Y} {\bm X}^H({\bm X}{\bm X}^H)^{-1}.
\end{equation}
Since each channel coefficient follows a $\mathcal{C}\mathcal{N}(0,\,1/(\lambda+1))$ distribution, the normalized mean square error (NMSE) can be formulated as 
\begin{equation}\label{eq:NMSE}
\begin{aligned}
\text{NMSE}&=\frac{\sigma^{2}_{v}N_r}{N_rN_t}\text{Tr}\left(({\bm X}{\bm X}^H)^{-1}\right)\\
&=\frac{\sigma^{2}_{v}}{N_t}\text{Tr}\left(({\bm X}{\bm X}^H)^{-1}\right),
\end{aligned}
\end{equation}
where 
\begin{equation}
\begin{aligned}
&\bm{X}\bm{X}^H\\
&= 
\begin{pmatrix}
{\bm C}^{(0)}\left({\bm C}^{(0)}\right)^H & {\bm C}^{(0)}\left({\bm C}^{(1)}\right)^H & \cdots & {\bm C}^{(0)}\left({\bm C}^{(\lambda)}\right)^H\\
{\bm C}^{(1)}\left({\bm C}^{(0)}\right)^H & {\bm C}^{(1)}\left({\bm C}^{(1)}\right)^H & \cdots & {\bm C}^{(1)}\left({\bm C}^{(\lambda)}\right)^H \\
\vdots & \vdots &\dots & \vdots \\
{\bm C}^{(\lambda)}\left({\bm C}^{(0)}\right)^H & {\bm C}^{(\lambda)}\left({\bm C}^{(1)}\right)^H  & \cdots & {\bm C}^{(\lambda)}\left({\bm C}^{(\lambda)}\right)^H
\end{pmatrix}
\end{aligned}
\end{equation}
according to (\ref{eq:training}) and (\ref{eq:circular_matrix}).
From \cite{CZCP-1st}, the minimum NMSE can be obtained if and only if 
\begin{equation}\label{eq:NMSE_criteria}
\begin{aligned}
\bm{X}\bm{X}^H=M \bm{I}_{N_t(\lambda+1)}.
\end{aligned}
\end{equation}
That is,
\begin{equation}
\begin{aligned}
&\bm{C}^{(u)}_{g}\left(\bm{C}^{(u')}_{k}\right)^H=
\begin{cases}
M, & 0\leq u=u'\leq \lambda,\\ 
   &  0\leq g=k\leq N_{t}-1;\\
0, & 0\leq u\neq u'\leq \lambda,\\
&  \text{or}~0\leq g\neq k\leq N_{t}-1,
\end{cases}
\end{aligned}
\end{equation}
implying 
\begin{equation}\label{eq:C2}
\theta(\bm{C}_{g},\bm{C}_{k};u)=
\begin{cases}
M,& g=k, u=0; \\
0,& g= k, 1\leq u\leq \lambda;\\
0, & g\neq k, 0\leq u\leq \lambda.\\
\end{cases}
\end{equation}
Therefore, by applying (\ref{eq:NMSE_criteria}) to (\ref{eq:NMSE}), the minimum NMSE is given as 
\begin{equation}
\text{minimum NMSE}=\frac{{\sigma_v^{2}(\lambda+1)}}{M}.
\end{equation}
\begin{lemma}\label{lemma:ZCZ_MSE}
According to (\ref{eq:C2}), one can attain the minimum NMSE for the channel estimation if and only if the SM training matrix is an $(N,L,Z,\mathcal{S})$-SZCZ array with $Z\geq \lambda$ and $\mathcal{S}=(L-M)/L$. 
\end{lemma}

As only one TA is activated during each time-slot, for the SM training matrix provided in (\ref{eq:training}), two design criteria are outlined below: 

{\it Criterion} (C1)
Each column in the training matrix, i.e, $(C_{0,i},C_{1,i},\ldots,C_{N-1,i})^T$ for $0\leq i\leq L-1$, contains only one non-zero entry. 

{\it Criterion} (C2)
The training matrix must meet the condition specified in (\ref{eq:C2}).

%

Since the SZCZ arrays have already fulfilled criterion (C2) according to {\it Lemma \ref{lemma:ZCZ_MSE}}, the primary objective is to design the SZCZ array to fulfill criterion (C1). 
\begin{defin}\label{defin:SZZ_matrix}
The SZCZ array that fulfills criterion (C1) is referred to as the {\it SZCZ training matrix}.
\end{defin}
\begin{rmk}\label{rmk:CZCP_training}
In \cite{CZCP-1st}, the CZCP-based training matrix is constructed by arranging CZCPs row by row. Specifically, for a training matrix of size $2^n\times 2^m$, each row consists of a CZCP of length $2^{m-n-1}$ and $2^{m}-2^{m-n}$ zero entries. If taking $N_t=4$ for example, the CZCP-based training matrix of size $4\times 2^m$ is given as follows:
\begin{equation*}\label{CZCPframe}
\left( \begin{array}{llll:llll}{\bm D}_0 & {\bm 0} & {\bm 0} & {\bm 0} & {\bm D}_1 & {\bm 0} & {\bm 0} & {\bm 0} \\ 
{\bm 0} & {\bm D}_0 & {\bm 0} & {\bm 0} & {\bm 0} & {\bm D}_1 & {\bm 0} & {\bm 0}\\ 
{\bm 0} & {\bm 0} & {\bm D}_0 & {\bm 0} & {\bm 0} & {\bm 0} & {\bm D}_1 & {\bm 0}\\ 
{\bm 0} & {\bm 0} & {\bm 0} & {\bm D}_0 & {\bm 0} & {\bm 0} & {\bm 0} & {\bm D}_1 \end{array}\right)_{4\times 2^m},
\end{equation*}
where $({\bm D}_0,{\bm D}_1)$ forms a $(2^{m-3},Z)$-CZCP and ${\bm 0}$ represents ${\bm 0}_{2^{m-3}}$.  It can be observed that any two rows are cyclically shifted versions of each other, ensuring that each column has only one non-zero entry, thereby fulfilling the design criteria for the SM training matrix. For different values of $N_t$, the same logic as provided above can be straightforwardly applied.
\end{rmk}

The training matrices currently in use follow the framework proposed by \cite{CZCP-1st} as stated in {\it Remark \ref{rmk:CZCP_training}}, relying on existing CZCPs or CZCSs as kernel sequences. These kernel sequences must follow specific patterns to construct sparse training matrices that meet criteria (C1) and (C2) in SM. In fact, they can be considered as specific instances of SZCZ training matrices. In contrast, this paper aims to directly develop general constructions of SZCZ training matrices with larger ZCZ widths, without relying on any specific sequences as kernels. In the following sections, we may use ``SZCZ matrix'' to denote SZCZ training matrix for simplicity.
\subsection{Proposed SZCZ Training Matrix for SM System}
By leveraging 2D RGBFs, we present direct constructions of SZCZ matrices for the SM system. Furthermore, we will show that the proposed SZCZ matrix exhibits a larger ZCZ width compared to existing CZCP-based and CZCS-based training matrices. First, we introduce the basic construction of SZCZ matrices in the following theorem.
\begin{thm}\label{thm:SZCZ_1}
Given positive integers $m$ and $n$, denote $\pi$ as the permutation of the set $\{1,2,\ldots,m\}$. Let ${\bm x}=(x_{\pi(m-n+1)},x_{\pi(m-n+2)},\ldots,x_{\pi(m)})$ and ${\bm y}=(y_1,y_2,\ldots,y_n)$. The 2D RGBF is
\begin{equation}\label{eq:SZCZ_thm1}
\begin{aligned}
f|_{\bm x=\bm y}=&\frac{q}{2}\sum_{l=1}^{m-n-1} x_{\pi(l)}x_{\pi(l+1)}+\sum_{l=1}^{m}\mu_{l}x_{l}+\sum_{s=1}^{n}\kappa_{s}y_{s}+\mu_{0}
\end{aligned}
\end{equation}
where $\mu_{l},\kappa_{s}\in \mathbb{Z}_q$, and $\mu_{m}\in \{0,q/2\}$. If $\pi(1)=m$ and 
$\pi(m-n+a)\in \{m-n,m-n+1,\ldots,m-1\}$ for $\alpha=1,2,\ldots,n$, then the matrix 
\begin{equation}
{\bm C=\bm f|_{\bm x=\bm y}}
\end{equation}
forms a $(2^n,2^m,2^{\pi(2)-1},\mathcal{S})$-SZCZ matrix where $\mathcal{S}=({2^n-1})/{2^n}$.
\end{thm}
\begin{IEEEproof}
The proof is given in Appendix \ref{apxA}
\end{IEEEproof}
\begin{rmk}
The $(2^n,2^m,2^{\pi(2)-1},\mathcal{S})$-SZCZ matrix constructed from {\it Theorem \ref{thm:SZCZ_1}} can be directly employed as a training matrix of size $2^n\times 2^m$ in the SM system since it fulfills criteria (C1) and (C2).
\end{rmk}
\begin{crly}\label{crly:SZCZ_CZCP}
Taking $\pi(2)=m-n-1$ and $\pi(m-n+\alpha)=m-n+\alpha-1$ for $\alpha=1,2,\ldots,n$ in {\it Theorem \ref{thm:SZCZ_1}}, the constructed $(2^n,2^m,2^{m-n-2},\mathcal{S})$-SZCZ matrix is reduced to the training matrix constructed from the perfect CZCPs of length $2^{m-n-1}$ in \cite{CZCP-1st}.
\end{crly}
\begin{IEEEproof}
The proof is given in Appendix \ref{apxB}. 
\end{IEEEproof}

Next, we use an example to illustrate that the proposed SZCZ matrix can be reduced to the CZCP-based training matrix.
\begin{eg}\label{eg:SZCZ_CZCP}
Considering $q=2$, $m=5$, and $n=2$, the constructed training matrix is of size $4\times 32$. Let $\pi=(5,2,1,3,4)$, $\bm x=(x_3,x_4)$, and $\bm y=(y_1,y_2)$. According to {\it Theorem \ref{thm:SZCZ_1}}, the 2D RGBF is 
\begin{equation}
f|_{\bms x=\bms y}=x_5x_2+x_2x_1
\end{equation}
where $\mu_{l}=0$ and $\kappa_{s}=0$ for all $l$ and $s$. The matrix  ${\bm C}={\bm f|_{\bm x=\bm y}}$ is a sparse $(4,32,2,3/4)$-SZCZ matrix given by (\ref{eq:CZCP_eg}). The PACFs of ${\bm C}_0$ and PCCFs of ${\bm C}_0$ and ${\bm C}_3$ are shown in Fig.~\ref{fig:SZCZ_CZCP} for shifts $u=0,1,\ldots,16$. Clearly, each column has only one non-zero entry and the ZCZ width is indeed $2$. It can be observed that the constructed SZCZ matrix is identical to the CZCP-based training matrix in \cite{CZCP-1st}, which is composed of a perfect $(4,2)$-CZCP $((+++-),(++-+))$. 
\begin{figure*}[ht!]
\begin{equation}\label{eq:CZCP_eg}
\begin{aligned}
\arraycolsep=2pt
{\bm C}=
\begin{pmatrix}
{\bm C}_0\\
{\bm C}_1\\
{\bm C}_{2}\\
{\bm C}_{3}
\end{pmatrix}
= \left( {\begin{array}{cccccccccccccccccccccccccccccccccccccccccccccccccccccccccccccccccccccccccccccccccccccccccccccccccccccccccccccccccccccccccccccccccc}
+&	+&	+&	-&	0&	0&	0&	0&	0&	0&	0&	0&	0&	0&	0&	0&	+&	+&	-&	+&	0&	0&	0&	0&	0&	0&	0&	0&	0&	0&	0&	0\\
0&	0&	0&	0&	+&	+&	+&	-&	0&	0&	0&	0&	0&	0&	0&	0&	0&	0&	0&	0&	+&	+&	-&	+&	0&	0&	0&	0&	0&	0&	0&	0\\
0&	0&	0&	0&	0&	0&	0&	0&	+&	+&	+&	-&	0&	0&	0&	0&	0&	0&	0&	0&	0&	0&	0&	0&	+&	+&	-&	+&	0&	0&	0&	0\\
0&	0&	0&	0&	0&	0&	0&	0&	0&	0&	0&	0&	+&	+&	+&	-&	0&	0&	0&	0&	0&	0&	0&	0&	0&	0&	0&	0&	+&	+&	-&	+\\
\end{array}}
\right).
\end{aligned}
\end{equation}
\end{figure*}
\begin{figure}[ht!]
	\centering
	\includegraphics[width = 3.5in]{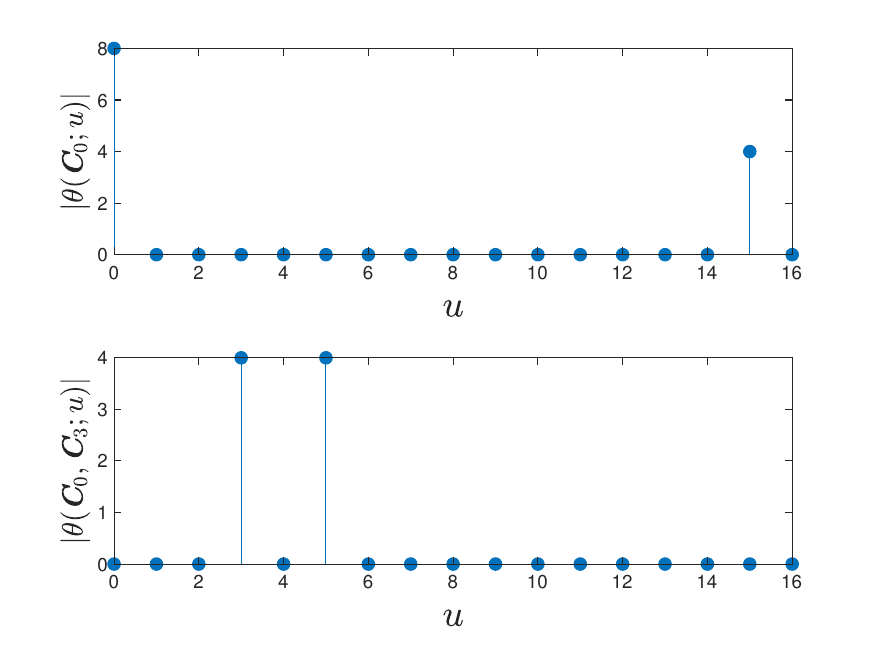}
	\caption{PACFs of ${\bm C}_0$ and PCCFs of  ${\bm C}_0$ and ${\bm C}_3$ in Example~ \ref{eg:SZCZ_CZCP}.}
    \label{fig:SZCZ_CZCP}	
\end{figure}
\end{eg}

While the $(2^n,2^m,2^{\pi(2)-1},\mathcal{S})$-SZCZ matrix proposed in {\it Theorem \ref{thm:SZCZ_1}} can be directly utilized as the training matrix in the SM system, its largest ZCZ width is limited to $2^{m-n-2}$ when $\pi(2)=m-n-1$. However, a larger ZCZ width is desirable for the SZCZ matrix since it enables greater tolerance for multipaths, as stated in {\it Lemma \ref{lemma:ZCZ_MSE}}. Therefore, we propose a second construction of SZCZ matrices with larger ZCZ widths in the following theorem.
\begin{thm}\label{thm:SZCZ}
For positive integers $m$ and $n$ with $m>n$, let nonempty set $I_1,I_2,\ldots I_n$ be a partition of $\{1,2,\ldots,m\}$, $\pi_{\alpha}$ be a bijection from $\{1,2,\ldots,m_{\alpha}\}$ to $I_{\alpha}$ for $\alpha=1,2,\ldots,n$ where $m_{\alpha}=|I_{\alpha}|$. Let ${\bm x}=(x_{\pi_{1}(m_1)},x_{\pi_{2}(m_2)},\ldots,x_{\pi_{n}(m_n)})$ and $\bm y=(y_{1},y_{2},\ldots,y_{n})$. The 2D RGBF is 
\begin{equation}\label{eq:SZCZ_thm2}
\begin{aligned}
f|_{\bm x=\bm y}=&\frac{q}{2}\sum_{\alpha=1}^{n}\sum_{\beta=1}^{m_{\alpha-1}}x_{\pi_{\alpha}(\beta)}x_{\pi_{\alpha}(\beta+1)}+x_{\pi_{\alpha}(m_{\alpha})}y_{\alpha}
\\&+\sum_{l=1}^{m}\mu_{l}x_{l}+\sum_{s=1}^{n}\kappa_{s}y_{s}+\mu_{0}
\end{aligned}
\end{equation}
where $\mu_{l},\kappa_{s}\in \mathbb{Z}_q$, and $\mu_{m}\in \{0,q/2\}$. If $\pi_{\alpha}(1)=m-\alpha+1$ for $1\leq \alpha\leq n$, the matrix
\begin{equation}
\bm C={\bm f}|_{{\bms x}={\bms y}}
\end{equation}
is a $(2^n,2^m,2^{\pi_{1}(2)-1},\mathcal{S})$-SZCZ matrix where $\mathcal{S}=(2^n-1)/2^n$.
\end{thm}
\begin{IEEEproof}
The proof is given in Appendix \ref{apxC}.
\end{IEEEproof}
\begin{rmk}
If taking $\pi_{1}(2)=m-n$ in {\it Theorem \ref{thm:SZCZ}}, we can obtain the SZCZ matrix of size $2^n \times 2^m$ with the largest ZCZ width $Z=2^{m-n-1}$. 
\end{rmk}

In the following example, we show that the proposed SZCZ matrix exhibits a larger ZCZ width compared to existing CZCP-based and CZCS-based training matrices.
\begin{eg}\label{eg:SZCZ}
For $q=2$, $m=6$, and $n=2$, denote $I_1=\{3,4,6\}$ and $I_2=\{2,1,5\}$ with $\pi_{1}=(6,4,3)$ and $\pi_{2}=(5,2,1)$ according to {\it Theorem \ref{thm:SZCZ}}. Let ${\bm x}=(x_3,x_1)$ and ${\bm y}=(y_1,y_2)$. The 2D RGBF is 
\[f|_{\bms x=\bms y}=x_6x_4+x_4x_3+x_3y_1+x_5x_2+x_2x_1+x_1y_2,\]
where $\mu_{l}=0$ and $\kappa_s=0$. The matrix 
\[{\bm C}={\bm f}|_{\bms x=\bms y}\]
is a $(4,64,8,3/4)$-SZCZ matrix given by (\ref{eq:SZCZ_eg}). The PACFs of ${\bm C}_0$ and PCCFs of ${\bm C}_0$ and ${\bm C}_3$ are depicted in Fig.~\ref{fig:SZCZ}. Clearly, each column has only one non-zero entry. The ZCZ width is indeed $8$ whereas the CZCP-based \cite{CZCP-1st} and CZCS-based \cite{Huang_22} training matrices of size $4\times 64$ have ZCZ widths of $4$ and $3$, respectively. 
\begin{figure*}[htb]
\begin{equation}\label{eq:SZCZ_eg}
\begin{aligned}
\arraycolsep=0.25pt
\footnotesize 
{\bm C}
= \left( {\begin{array}{cccccccccccccccccccccccccccccccccccccccccccccccccccccccccccccccccccccccccccccccccccccccccccccccccccccccccccccccccccccccccccccccccc}
+&	0&	+&	0&	0&	0&	0&	0&	+&	0&	+&	0&	0&	0&	0&	0&	+&	0&	-&	0&	0&	0&	0&	0&	+&	0&	-&	0&	0&	0&	0&	0&	+&	0&	+&	0&	0&	0&	0&	0&	-&	0&	-&	0&	0&	0&	0&	0&	+&	0&	-&	0&	0&	0&	0&	0&	-&	0&	+&	0&	0&	0&	0&	0\\
0&	0&	0&	0&	-&	0&	-&	0&	0&	0&	0&	0&	+&	0&	+&	0&	0&	0&	0&	0&	-&	0&	+&	0&	0&	0&	0&	0&	+&	0&	-&	0&	0&	0&	0&	0&	-&	0&	-&	0&	0&	0&	0&	0&	-&	0&	-&	0&	0&	0&	0&	0&	-&	0&	+&	0&	0&	0&	0&	0&	-&	0&	+&	0\\
0&	-&	0&	+&	0&	0&	0&	0&	0&	-&	0&	+&	0&	0&	0&	0&	0&	-&	0&	-&	0&	0&	0&	0&	0&	-&	0&	-&	0&	0&	0&	0&	0&	-&	0&	+&	0&	0&	0&	0&	0&	+&	0&	-&	0&	0&	0&	0&	0&	-&	0&	-&	0&	0&	0&	0&	0&	+&	0&	+&	0&	0&	0&	0\\
0&	0&	0&	0&	0&	+&	0&	-&	0&	0&	0&	0&	0&	-&	0&	+&	0&	0&	0&	0&	0&	+&	0&	+&	0&	0&	0&	0&	0&	-&	0&	-&	0&	0&	0&	0&	0&	+&	0&	-&	0&	0&	0&	0&	0&	+&	0&	-&	0&	0&	0&	0&	0&	+&	0&	+&	0&	0&	0&	0&	0&	+&	0&	+\\
\end{array}}
\right).
\end{aligned}
\end{equation}
\end{figure*}
\begin{figure}[htbp]
	\centering
	\includegraphics[width = 3.5in]{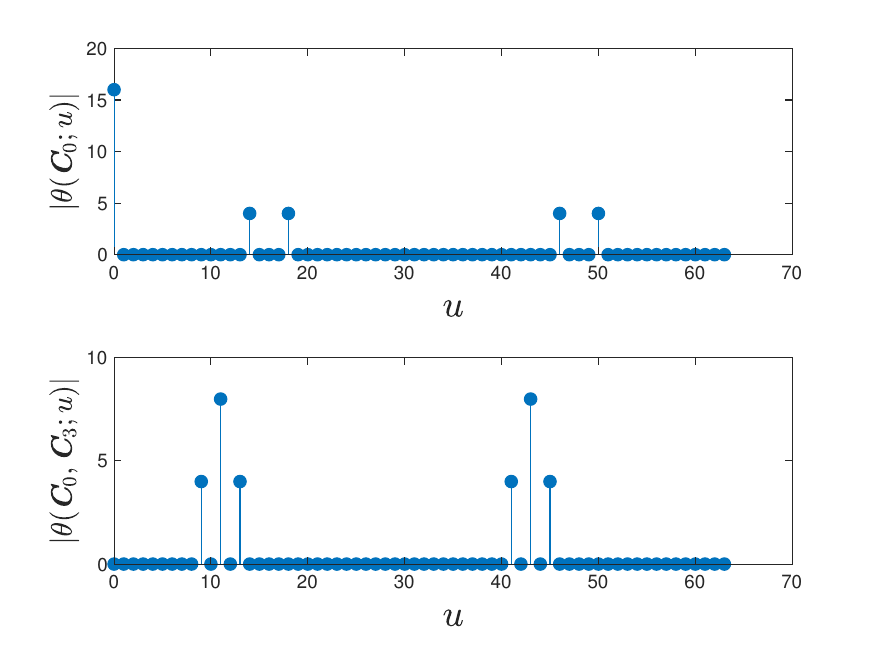}
	\caption{PACFs of ${\bm C}_0$ and PCCFs of  ${\bm C}_0$ and ${\bm C}_3$ in Example~ \ref{eg:SZCZ}.}
    \label{fig:SZCZ}	
\end{figure}
\end{eg}

\begin{rmk}
The proposed SZCZ matrices from {\it Theorem \ref{thm:SZCZ_1}} and {\it Theorem \ref{thm:SZCZ}} can be directly utilized as the training matrices in the SM system. In contrast, existing SM training schemes primarily follow the training framework proposed by \cite{CZCP-1st}, where CZCPs (or CZCSs) are employed as kernel sequences and must be arranged in a specific pattern to generate sparse training matrices for the SM system.
\end{rmk}

In Table \ref{table}, taking the same sparsity and a training matrix of size $2^{n}\times 2^{m}$, we compare the proposed SZCZ matrix with the existing CZCP-based and CZCS-based training matrices. Since the training matrix is of size $2^{n}\times 2^{m}$, we use perfect CZCPs \cite{CZCP-1st} and CZCSs \cite{Huang_22} of power-of-two lengths for comparison. Clearly, our proposed training design exhibits larger ZCZ widths, indicating greater tolerance for delay spread. This is because the types of training matrices proposed by \cite{CZCP-1st} and \cite{Huang_22} follow specific patterns, requiring the utilization of existing CZCPs and CZCSs, respectively, to construct sparse training matrices. As a consequence, the ZCZ widths of these training matrices are constrained by the properties of the CZCPs or CZCSs. On the other hand, our proposed SZCZ matrices based on 2D RGBFs can be directly used as training matrices in the SM system without the help of any kernel CZCPs or CZCSs, thereby offering greater flexibility in designing SM training matrices to obtain larger ZCZ widths.
\begin{table*}[htb]
\centering
\caption{Comparison of Different Frameworks for SM Training Matrices of Size $2^{n}\times 2^{m}$}\label{table}
\begin{tabular}{|c||c|c|c|c|c|c|}\hline
Method                                               & Largest ZCZ width                & Based on                      & Note                                   \\ \hline
\cite{CZCP-1st}                                   &   $2^{m-n-2}$         &  CZCP of length $2^{m-n-1}$  &        \\ \hline
\cite{Huang_22}                                      & $2^{m-n-k}-1$         & CZCS of set size $2^k$ and of length $2^{m-n-k}$  & $k\geq 2$      \\ \hline
{\it Theorem \ref{thm:SZCZ}}          & $2^{m-n-1}$ & 2D RGBF  & $m>n$
\\ \hline
\end{tabular}
\end{table*}

\section{Simulation Results}\label{sec:Simulation}
\begin{figure*}[htb]
 \begin{subfigure}{.5\textwidth}
 \parbox[][7cm][c]{\linewidth}{
         \centering
         \includegraphics[width=9cm]{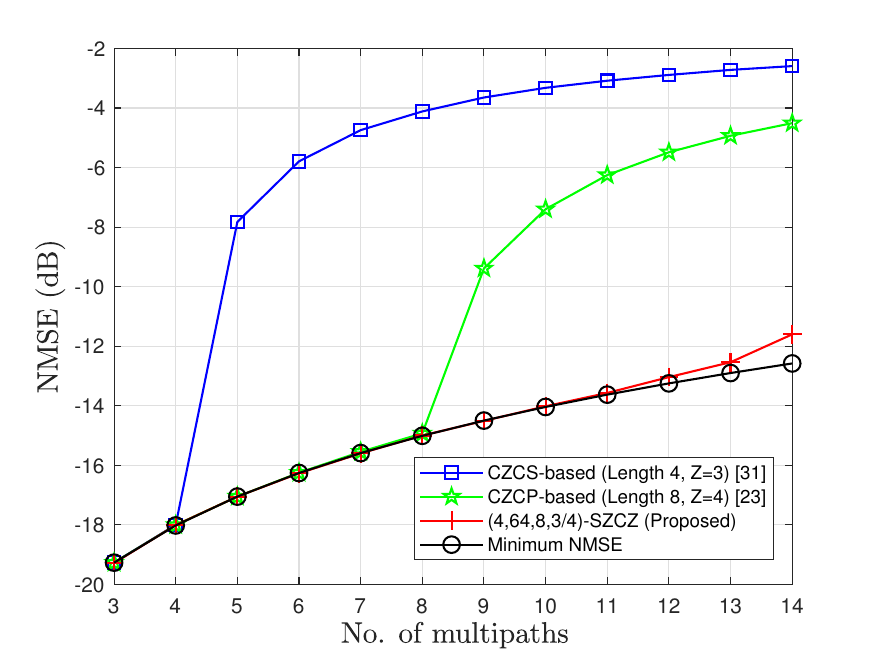}}
         \caption{SNR is $12$ dB.}
         \label{fig:MSE_path}
 \end{subfigure}
 \begin{subfigure}{.5\textwidth}
         \centering
         \includegraphics[width=9.1cm]{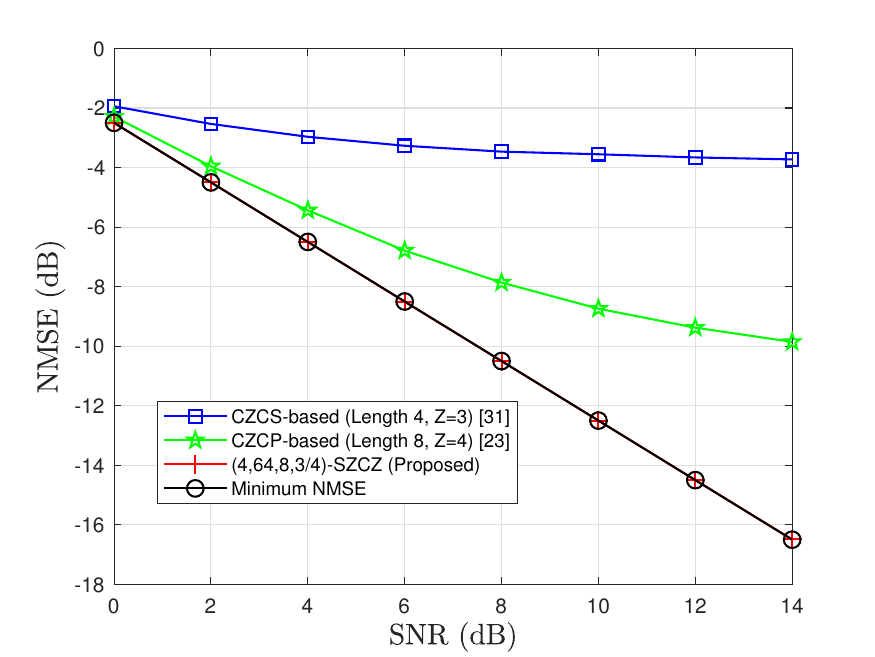}
         \caption{No. of multipaths is $9$.}
         \label{fig:MSE_SNR}
 \end{subfigure}
 \caption{Comparison of NMSE performance for different training matrices with $N_t=4$ and $N_r=4$.}\label{MSE_SZCZ}
\end{figure*}

\begin{figure}[htb]
	\centering
	\includegraphics[width = 3.5in]{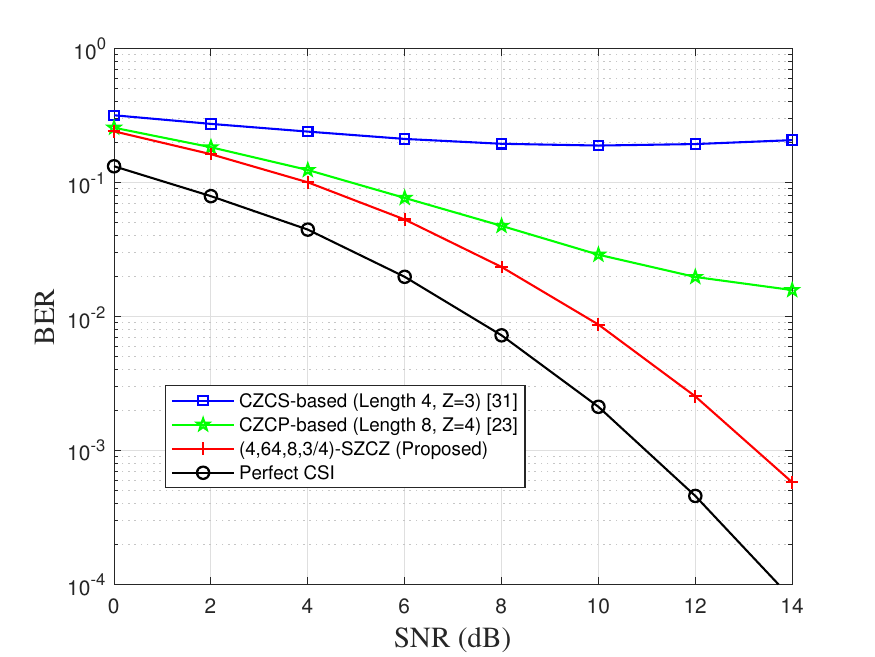}
	\caption{Comparison of BER performance for different training schemes with $9$ multipaths.}
    \label{fig:BER_SZCZ}	
\end{figure}

In this section, we evaluate the channel estimation performance of the proposed SZCZ matrix over quasi-static frequency-selective channels with $(\lambda+1)$-multipaths, as described in (\ref{eq:channel}). Throughout the simulations, we set $N_t=4$, $N_r=4$, and the training matrix is of size $4\times 64$ with elements drawn from the alphabet $\{+1,0,-1\}$. Let us consider the constructed $(4,64,8,3/4)$-SZCZ matrix as given in (\ref{eq:SZCZ_eg}). For comparison, we utilize a perfect CZCP of length $8$ and a CZCS with set size of $4$ and length $4$ to construct $4\times 64$ training matrices. The CZCP-based training matrix \cite{CZCP-1st} is given by 
\begin{equation}\label{CZCPframe}
\left( \begin{array}{llll:llll}{\bm D}_0 & {\bm 0} & {\bm 0} & {\bm 0} & {\bm D}_1 & {\bm 0} & {\bm 0} & {\bm 0} \\ 
{\bm 0} & {\bm D}_0 & {\bm 0} & {\bm 0} & {\bm 0} & {\bm D}_1 & {\bm 0} & {\bm 0}\\ 
{\bm 0} & {\bm 0} & {\bm D}_0 & {\bm 0} & {\bm 0} & {\bm 0} & {\bm D}_1 & {\bm 0}\\ 
{\bm 0} & {\bm 0} & {\bm 0} & {\bm D}_0 & {\bm 0} & {\bm 0} & {\bm 0} & {\bm D}_1 \end{array}\right)_{4\times 64},
\end{equation}
where ${\bm 0}$ represents ${\bm 0}_8$ and $({\bm D}_0,{\bm D}_1)=$ $(+++-++-+,+++---+-)$ is the perfect $(8,4)$-CZCP. The CZCS-based training matrix \cite{Huang_22} is given by
\begin{equation}
\begin{aligned}
\arraycolsep=1.5pt
\left( {\begin{array}{llll:llll:llll:llll}
  \bm s_0 & \bm 0 & \bm 0 & \bm 0 &  \bm s_1 & \bm 0 & \bm 0 & \bm 0 & \bm s_2 & \bm 0 & \bm 0 & \bm 0 & \bm s_3 & \bm 0 & \bm 0 & \bm 0 \\
  \bm 0 & \bm s_0 & \bm 0 & \bm 0 &  \bm 0 & \bm s_1 & \bm 0 & \bm 0 & \bm 0 & \bm s_2 & \bm 0 & \bm 0 & \bm 0 & \bm s_3 & \bm 0 & \bm 0 \\
  \bm 0 & \bm 0 & \bm s_0 & \bm 0 & \bm 0 & \bm 0 &  \bm s_1 & \bm 0 &  \bm 0 & \bm 0 & \bm s_2 & \bm 0  & \bm 0 & \bm 0 & \bm s_3 & \bm 0 \\
  \bm 0 & \bm 0 & \bm 0 & \bm s_0 & \bm 0 & \bm 0 & \bm 0 & \bm s_1  & \bm 0 & \bm 0 & \bm 0 & \bm s_2 & \bm 0 & \bm 0 & \bm 0 & \bm s_3
\end{array}}\right)_{4\times 64},\label{CZCSframe}
\end{aligned}
\end{equation}
where $({\bm s}_0,{\bm s}_1,{\bm s}_2,{\bm s}_3)=$ $(+-++,+++-,+-++,---+)$ forms the CZCS of length 4 and ${\bm 0}$ denotes ${\bm 0}_4$. It can be verified that the ZCZ widths of CZCP-based and CZCS-based training matrices from (\ref{CZCPframe}) and (\ref{CZCSframe}) are $4$ and $3$, respectively, as illustrated in Table \ref{table}. Additionally, we can observe that several CZCPs and CZCSs need to be arranged in specific patterns, respectively, to generate regular sparse matrices. The NMSE performances of channel estimation using different training matrices are depicted in Fig. \ref{MSE_SZCZ}. In Fig.\mbox{\ref{MSE_SZCZ}-\subref{fig:MSE_path}}, we compare NMSE performances with varying numbers of multipaths at SNR of $12$ dB. The NMSE performances of CZCP-based and CZCS-based training matrices degrade when the delay spread exceeds their respective ZCZ widths. Additionally, Fig.\mbox{\ref{MSE_SZCZ}-\subref{fig:MSE_SNR}} provides NMSE performances with varying SNRs when the number of multipaths is $9$. We can observe that the NMSE performance of the proposed $(4,64,3/4)$-SZCZ matrix aligns with the curve of the minimum NMSE. However, the NMSE performances of CZCP-based and CZCS-based training matrices degrade when the delay spread exceeds their respective ZCZ widths. It is noteworthy that the NMSE of our proposed SZCZ matrix with ZCZ width $Z=8$ achieves the minimum NMSE when the number of multipaths is 9 or fewer. Subsequently, using minimum MSE (MMSE) equalizer \cite{SM_ZPSC}, we analyze the bit error rate (BER) performances of different training schemes with $9$ multipaths in Fig. \ref{fig:BER_SZCZ}. Clearly, the BER performance of our proposed SZCZ-based training scheme outperforms the CZCP-based and CZCS-based training schemes and is closest to the perfect CSI curve. Compared to the perfect CSI curve, the approximately 2 dB degradation in the BER curve of our SZCZ-based training scheme is due to channel estimation errors from the LS estimator.
\section{Conclusion}\label{sec:conclusion}
In this paper, we have proposed a novel training matrix design for SM systems. First, we have introduced a new class of 2D arrays, i.e., the SZCZ array, which consists of several zero entries. Any two rows of an SZCZ array possess the zero periodic auto- and cross-correlation zone property. By fully leveraging the sparsity and correlation properties of SZCZ arrays, we have shown that certain SZCZ arrays, where each column has only one non-zero entry, can be used as training matrices for SM systems, referred to as SZCZ training matrices. Furthermore, we have introduced the concept of 2D RGBFs and proposed direct constructions of SZCZ training matrices with large ZCZ widths and controllable sparsity levels in {\it Theorem \ref{thm:SZCZ_1}} and {\it Theorem \ref{thm:SZCZ}}. Note that CZCP-based training matrices can be regarded as the special case of the proposed training matrices as stated in {\it Corollary \ref{crly:SZCZ_CZCP}}. Compared with existing CZCP-based and CZCS-based training matrices, the proposed SZCZ matrices exhibit larger ZCZ widths, indicating greater tolerance for delay spread, and do not require any CZCPs or CZCSs as kernel sequences. The simulation results indicate that the proposed SZCZ-based training design outperforms the performance of existing alternatives in channel estimation over frequency-selective fading channels.
\begin{appendices}
\section{Proof of Theorem \ref{thm:SZCZ_1}}\label{apxA}
\begin{IEEEproof}
For the SZCZ matrix 
\[{\bm C}=({\bm C}_0^T,{\bm C}_1^T,\ldots,{\bm C}_{2^{n}-1}^T)^T\]
constructed from {\it Theorem \ref{thm:SZCZ_1}}, where
\begin{equation}\label{eq:SZCZ_matrix}
\begin{pmatrix}
{\bm C}_0\\
{\bm C}_1\\
\vdots \\
{\bm C}_{2^n-1}
\end{pmatrix}=
\begin{pmatrix}
C_{0,0}& C_{0,1} & \cdots & C_{0,2^m-1}\\
C_{1,0} & C_{1,1} & \cdots & C_{1,2^m-1}\\
\vdots & \vdots & \ddots & \vdots   \\
C_{2^n-1,0} & C_{2^n-1,1} & \cdots & C_{2^n-1,2^m-1}
\end{pmatrix},
\end{equation}
it is necessary to demonstrate that criteria (C1) and (C2) are satisfied. 

In the first part, we commence by verifying (C1). Denote the $i$-th column sequence of the SZCZ matrix $\bm C$ as ${\bm C}^T_i=(C_{0,i},C_{1,i},\ldots,C_{2^n-1,i})^T$ where $0\leq i\leq 2^m-1$. Based on the concept of 2D RGBF, we have $C_{g,i}=\xi^{c_{g,i}}$ if $i_{\pi(m-n+\alpha)}=g_{\alpha}$ for $\alpha=1,2,\ldots,n$ or $C_{g,i}=0$ if $i_{\pi(m-n+\alpha)}\neq g_{\alpha}$ for some $1\leq \alpha\leq n$. For the given $i$, since
\begin{equation}
i=\sum_{l=1}^{m-n}i_{\pi(l)}2^{\pi(l)-1}+\sum_{\alpha=1}^{n}i_{\pi(m-n+\alpha)}2^{\pi(m-n+\alpha)-1},
\end{equation}
there exists only an integer $\hat{g}$ with 
\begin{equation}
\hat{g}=\sum_{\alpha=1}^{n}\hat{g}_{\alpha}2^{\alpha-1}=\sum_{\alpha=1}^{n}i_{\pi(m-n+\alpha)}2^{\alpha-1},
\end{equation}
where $0\leq \hat{g}\leq 2^{n}-1$. Hence, for the $i$-th column sequence ${\bm C}^T_i$, we have $C_{\hat{g},i}= \xi^{c_{\hat{g},i}}$ and $C_{g,i}=0$ for $g\neq \hat{g}$, implying (C1) holds.

Next, we examine (C2). Specifically,
\begin{equation}
\begin{aligned}
{\theta}(\bm C_{g}, \bm C_{k}; u)&=\sum_{i=0}^{L-1}C_{g,(i+u)_L}C_{k,i}^{*}\\
&=\begin{cases}
0, & g= k, 1\leq u\leq Z,\\
0& g\neq k, 0\leq u\leq Z,
\end{cases}
\end{aligned}
\end{equation} 
where $L=2^m$ and $Z=2^{\pi(2)-1}$. Denote $j=(i+u)_L$ and let $(i_1,i_2,\ldots,i_m)$ and $(j_1,j_2,\ldots,j_m)$ represent binary vectors of $i$ and $j$, respectively. Besides, we also let $(g_1,g_2,\ldots,g_n)$ and $(k_1,k_2,\ldots,k_n)$ be binary vectors of $g$ and $k$, respectively. We have ${\bm C}_g=(C_{g,0},C_{g,1},\ldots,C_{g,2^{m}-1})$ where $C_{g,i}=\xi^{c_{g,i}}$ if $i_{\pi(m-n+\alpha)}=g_{\alpha}$ for $\alpha=1,2,\ldots,n$ or $C_{g,i}=0$ if $i_{\pi(m-n+\alpha)}\neq g_{\alpha}$ for some $1\leq \alpha\leq n$. Therefore, we only consider the multiplications involving non-zero entries, i.e., $C_{g,j}C_{k,i}^{*}\neq 0$, implying $j_{\pi(m-n+\alpha)}=g_{\alpha}$ and $i_{\pi(m-n+\alpha)}=k_{\alpha}$ for $\alpha=1,2,\ldots,n$. In what follows, three cases are taken into account. 

{\it Case 1}: $u\neq 0$ and $i_m= j_m$. In this case, we assume that $v$ is the smallest integer such that $i_{\pi(v)}\neq j_{\pi(v)}$. Then we let $i'$ and $j'$ be integers distinct from $i$ and $j$, respectively, in the position $\pi(v-1)$, i.e., $i'_{\pi(v-1)}=1-i_{\pi(v-1)}$ and $j'_{\pi(v-1)}=1-j_{\pi(v-1)}$. Then we have
\begin{equation}
 \begin{aligned}
&c_{g,j}-c_{g,j'}\\
&=\frac{q}{2}(j_{\pi(v-2)}j_{\pi(v-1)}-j_{\pi(v-2)}j'_{\pi(v-1)}\\
&+j_{\pi(v-1)}j_{\pi(v)}-j'_{\pi(v-1)}j_{\pi(v)})\\
&+\mu_{\pi(v-1)}j_{\pi(v-1)}-\mu_{\pi(v-1)}j'_{\pi(v-1)}\\
&\equiv \frac{q}{2}(j_{\pi(v-2)}+j_{\pi(v)})+\mu_{\pi(v-1)}(2j_{\pi(v-1)}-1)\pmod q. 
\end{aligned}
\end{equation}
Likewise, 
\begin{equation}
 \begin{aligned}
&c_{k,i'}-c_{k,i}\equiv \frac{q}{2}(i_{\pi(v-2)}+i_{\pi(v)})+\mu_{\pi(v-1)}(1-2i_{\pi(v-1)})\\
&\qquad\qquad\qquad\qquad\qquad\qquad\qquad\qquad\qquad\quad \pmod q. 
\end{aligned}
\end{equation}
Since $j_{\pi(v-1)}=i_{\pi(v-1)}$ and $j_{\pi(v-2)}=i_{\pi(v-2)}$, we obtain
\begin{equation}
c_{g,j}-c_{k,i}-c_{g,j'}+c_{k,i'}\equiv \frac{q}{2}(j_{\pi(v)}-i_{\pi(v)})\equiv\frac{q}{2},
\end{equation}
which means $\xi^{c_{g,j}-c_{k,i}}+\xi^{c_{g,j'}-c_{k,i'}}=0$.

{\it Case 2}: $u\neq 0$ and $i_m\neq j_m$. Here, we have $i_{\pi(2)}\neq j_{\pi(2)}$. Suppose not, assume that $i_{\pi(2)}= j_{\pi(2)}$. Without loss of generality, we consider $j_m=1$ and $i_m=0$ and hence we obtain
 \begin{equation}
 \begin{aligned}
u=j-i & =  2^{m-1} + \sum_{l=1,l\neq \pi(2)}^{m-1}(j_l-i_l)2^{l-1} \\
 &\geq  2^{m-1} - \sum_{l=1}^{m-1}2^{l-1} + 2^{\pi(2)-1} = 2^{\pi(2)-1}+1
 \end{aligned}
\end{equation}
which contradicts the assumption that $u\leq 2^{\pi(2)-1}$. Therefore, for $i_{\pi(2)}\neq j_{\pi(2)}$, let $i'$ and $j'$ be integers distinct from $i$ and $j$, respectively, at $\pi(1)=m$, i.e., $i'_{m}=1-i_{m}$ and $j'_{m}=1-j_{m}$. Hence, it leads to
\begin{equation}
 \begin{aligned}
c_{g,j}-c_{g,j'}&=\frac{q}{2}\left(j_{m}{j}_{\pi(2)} - j'_{m}{j}_{\pi(2)} \right) + \mu_{m}j_{m} - \mu_{m}j'_{m}\\
&\equiv \frac{q}{2}{j}_{\pi(2)} + \mu_{m}\pmod{q},
\end{aligned}
\end{equation}
where $\mu_{m}\in \{0,q/2\}$. Due to the fact that $j_{\pi(2)}\neq i_{\pi(2)}$, we obtain $\xi^{c_{g,j}-c_{k,i}}+\xi^{c_{g,j'}-c_{k,i'}}=0$.

{\it Case 3}: $u=0$ and $g\neq k$. By invoking the proof provided in the first part, we possess $C_{g,i}C^{*}_{k,i}=0$ for $0\leq i\leq 2^{m}-1$, resulting in ${\theta}(\bm C_{g}, \bm C_{k}; 0)=0$.

From {Case 1} to {Case 3}, we validate that (C2) holds, thereby completing the proof.
\end{IEEEproof}
\section{Proof of Corollary \ref{crly:SZCZ_CZCP}}\label{apxB}
Before proceeding, we introduce the following lemma, which is useful in the proof of {\it Corollary \ref{crly:SZCZ_CZCP}}.
\begin{lemma}\label{lemma:CZCP}\cite{CZCP-1st}
For a positive integer $m'$, let $\pi$ be the permutation of the set $\{1,2,\ldots,m'\}$ with $\pi(1)=m'$. The GBF is $f=\frac{q}{2}\sum_{l=1}^{m'-1} x_{\pi(l)}x_{\pi(l+1)}+\sum_{l=1}^{m'}\mu_{l}x_{l}+\mu_{0},$
where $\mu_{l}\in \mathbb{Z}_q$. The pair 
\begin{equation}\label{eq:CZCP_GBF}
(\bm f,{\bm f}')=\left(\bm f,{\bm f}+\frac{q}{2}{\bm x}_{m'}\right)
\end{equation}
forms a perfect $q$-ary $(2^{m'},2^{m'-1})$-CZCP.
\end{lemma} 
\begin{IEEEproof}
We aim to demonstrate that by taking $\pi(2)=m-n-1$ and $\pi(m-n+\alpha)=m-n+\alpha-1$ for $\alpha=1,2,\ldots,n$ in {\it Theorem \ref{thm:SZCZ_1}}, any row of ${\bm C}$, i.e., ${\bm C}_g$ for $0\leq g\leq 2^n-1$, comprises a CZCP of length $2^{m-n-1}$, and $2^{m}-2^{m-n}$ zero entries as stated in {\it Remark \ref{rmk:CZCP_training}}. The $g$-th row of the SZCZ matrix ${\bm C}$ is given by ${\bm C}_g=(C_{g,0},C_{g,1},\ldots,C_{g,2^{m}-1})$, $0\leq g \leq 2^n-1$, where $C_{g,i}=\xi^{c_{g,i}}$ for $(i_{m-n},i_{m-n+1},\ldots,m-1)=(g_1,g_2,\ldots,g_n)$ or $C_{g,i}=0$ for $(i_{m-n},i_{m-n+1},\ldots,m-1)\neq (g_1,g_2,\ldots,g_n)$. 
We first consider the case for ${\bm C}_0$, which can be represented as
\begin{equation}
{\bm C}_0=({\bm D}_0,{\bm 0},{\bm D}_1,{\bm 0})_{1\times 2^m},
\end{equation}
where 
\begin{equation}
\begin{aligned}
{\bm D}_0&=(\xi^{{\bm d}_0})=(\xi^{c_{0,0}},\xi^{c_{0,1}},\ldots,\xi^{c_{0,2^{m-n-1}-1}}),\\
{\bm D}_1&=(\xi^{{\bm d}_1})=(\xi^{c_{0,2^{m-1}}},\ldots,\xi^{c_{0,2^{m-n-1}+2^{m-1}-1}}),
\end{aligned}
\end{equation}
and ${\bm 0}$ is a zero vector of length $2^{m-1}-2^{m-n-1}$. Since 
\begin{equation*}
\{\pi(2),\pi(3),\ldots,\pi(m-n)\}=\{1,2,\ldots,m-n-1\},
\end{equation*}
with $\pi(2)=m-n-1$, let $\pi_1(l)=\pi(l+1)$, $l=1,2,\ldots,m-n-1$, for ease of presentation. Hence, from (\ref{eq:SZCZ_thm1}), ${\bm d}_0$ and ${\bm d}_1$ can be expressed as
\begin{equation}\label{eq:SZCZ_CZCP}
\begin{aligned}
{\bm d}_0&=\frac{q}{2}\sum_{l=1}^{m-n-2} {\bm x}_{\pi_1(l)}{\bm x}_{\pi_1(l+1)}+\sum_{l=1}^{m-n-1}\mu_{l}{\bm x}_{l}+\mu_{0}{\bm 1},\\
{\bm d}_1&={\bm d}_0+\frac{q}{2}{\bm x}_{m-n-1},
\end{aligned}
\end{equation}
for the reason that $\pi_{1}(1)=\pi(2)=m-n-1$. It can be seen that the pair $({\bm d}_0,{\bm d}_1)$ takes the same form as in (\ref{eq:CZCP_GBF}) by setting $m'=m-n-1$ in {\it Lemma \ref{lemma:CZCP}}. Therefore, $({\bm d}_0,{\bm d}_1)$ is a perfect $q$-ary $(2^{m-n-1},2^{m-n-2})$-CZCP. Following a similar line of reasoning, we can derive that any ${\bm C}_g$ is composed of a perfect CZCP $({\bm d}_0,{\bm d}_1)$ and $2^{m}-2^{m-n}$ zero entries, thereby completing the proof.
\end{IEEEproof}
\section{Proof of Theorem \ref{thm:SZCZ}}\label{apxC}
\begin{IEEEproof}
According to {\it Theorem \ref{thm:SZCZ}}, we know $C_{g,i}=\xi^{c_{g,i}}$ if $i_{\pi_{\alpha}(m_{\alpha})}=g_{\alpha}$ for $\alpha=1,2,\ldots,n$ or $C_{g,i}=0$ if $(i_{\pi_{1}(m_1)},i_{\pi_{2}(m_2)},\ldots,i_{\pi_{n}(m_n)})\neq (g_1,g_2,\ldots,g_n)$.
Initially, we demonstrate that the criterion (C1) holds. Similarly, denote the $i$-th column sequence of the SZCZ matrix $\bm C$ as ${\bm C}^T_i=(C_{0,i},C_{1,i},\ldots,C_{2^n-1,i})^T$ where $0\leq i\leq 2^m-1$. For the given $i$, it can be written as 
\begin{equation}
i=\sum_{\alpha=1}^{n}\sum_{\beta=1}^{m_{\alpha-1}}i_{\pi_{\alpha}(\beta)}2^{\pi_{\alpha}(\beta)-1}+\sum_{\alpha=1}^{n}i_{\pi_{\alpha}(m_{\alpha})}2^{\pi_{\alpha}(m_{\alpha})-1}.
\end{equation}
Since $i_{\pi_{\alpha}(m_{\alpha})}=g_{\alpha}$ for all $\alpha=1,2,\ldots,n$, there exists only an integer $\hat{g}$ satisfying
\begin{equation}
\hat{g}=\sum_{\alpha=1}^{n}\hat{g}_{\alpha}2^{n-1}=\sum_{\alpha=1}^{n}i_{\pi_{\alpha}(m_{\alpha})}2^{\alpha-1},~0\leq \hat{g}\leq 2^{n}-1.
\end{equation}
Therefore, we have $C_{\hat{g},i}=\xi^{c_{\hat{g},i}}$ and $C_{g,i}=0$ for $g\neq \hat{g}$ in the $i$-th column sequence ${\bm C}^T_i$ which means (C1) holds.

Subsequently, we check (C2). That is, 
\begin{equation}
\begin{aligned}
{\theta}(\bm C_{g}, \bm C_{k}; u)&=\sum_{i=0}^{L-1}C_{g,(i+u)_L}C_{k,i}^{*}\\
&=\begin{cases}
0, & g= k, 1\leq u\leq Z,\\
0& g\neq k, 0\leq u\leq Z,
\end{cases}
\end{aligned}
\end{equation} 
where $L=2^m$. Similarly, let $j=(i+u)_L$ and let $(i_1,i_2,\ldots,i_m)$, $(j_1,j_2,\ldots,j_m)$, $(g_1,g_2,\ldots,g_n)$, and $(k_1,k_2,\ldots,k_n)$  represent binary vectors of $i$, $j$, $g$, and $k$, respectively. We only consider the multiplications involving non-zero entries, i.e., $C_{g,j}C_{k,i}^{*}\neq 0$, implying $j_{\pi_{\alpha}(m_{\alpha})}=g_{\alpha}$ and $i_{\pi_{\alpha}(m_{\alpha})}=k_{\alpha}$ for $\alpha=1,2,\ldots,n$. Three cases are considered below. 

{\it Case 1}: $u\neq 0$ and $j_{\pi_{1}(2)}=i_{\pi_{1}(2)}$. In this case, we will show that $j_{\pi_{\alpha}(1)}=i_{\pi_{\alpha}(1)}$ for all $\alpha=1,2,\ldots, n$. If we assume that $\gamma$ is the smallest integer such that $j_{\pi_{\gamma}(1)} \neq i_{\pi_{\gamma}(1)}$. Without loss of generality, we examine $j_{\pi_{\gamma}(1)}>i_{\pi_{\gamma}(1)}=0$, which implies $j_{m-\gamma+1}=1$ and $i_{m-\gamma+1}=0$. It leads to 
 \begin{equation}
 \begin{aligned}
u=j-i &=   2^{m-\gamma} + \sum_{l=1,l\neq \pi_1(2)}^{m-\gamma}(j_l-i_l)2^{l-1} \\
 &\geq  2^{m-\gamma} - \sum_{l=1}^{m-\gamma}2^{l-1} + 2^{\pi_1(2)-1} =  2^{\pi_1(2)-1}+1,
 \end{aligned}
\end{equation}
thereby contradicting the assumption of $u\leq 2^{\pi_1(2)-1}$.  Hence, we have $j_{\pi_{\alpha}(1)}=i_{\pi_{\alpha}(1)}$ for $1 \leq \alpha \leq n$. Let $\hat{\alpha}$ be the largest integer such that $j_{\pi_\alpha(\beta)} = i_{\pi_\alpha(\beta)}$ for $\alpha = 1,2,\ldots,\hat{\alpha}-1$ and $\beta = 1,2,\ldots,m_\alpha$. Let $v$ be the smallest integer with $j_{\pi_{\hat{\alpha}}(v)} \neq i_{\pi_{\hat{\alpha}}(v)}$. Denote $j'$ and $i'$ as integers distinct from $j$ and $i$, respectively, in the position $\pi_{\hat{\alpha}}(v-1)$. Following the similar arguments as provided in Case 1 of Appendix \ref{apxA}, we have
\begin{equation}
c_{g,j}-c_{k,i}-c_{g,j'}+c_{k,i'}\equiv \frac{q}{2}(j_{\pi_{\hat{\alpha}}(v)}-i_{\pi_{\hat{\alpha}}(v)})\equiv\frac{q}{2},
\end{equation}
indicating $\xi^{c_{g,j}-c_{k,i}}+\xi^{c_{g,j'}-c_{k,i'}}=0$.

{\it Case 2}: $u\neq 0$ and $j_{\pi_{1}(2)}\neq i_{\pi_{1}(2)}$. Likewise, let $j'$ and $i'$ be integers distinct from $j$ and $i$, respectively, i.e., $j'_{m}=1-j_m$ and $i'_{m}=1-i_m$. By using the similar derivation as given in Case 2 of Appendix \ref{apxA}, we obtain $\xi^{c_{g,j}-c_{k,i}}+\xi^{c_{g,j'}-c_{k,i'}}=0$.

{\it Case 3}: $u=0$ and $g\neq k$. By invoking the proof provided in the first part, we can derive $C_{g,i}C^{*}_{k,i}=0$ for $0\leq i\leq 2^{m}-1$, resulting in ${\theta}(\bm C_{g}, \bm C_{k}; 0)=0$.

Combining {Case 1} through {Case 3}, we confirm that (C2) holds, thus completing the proof.
\end{IEEEproof}
\end{appendices}
\bibliographystyle{IEEEtran}
\bibliography{IEEEabrv,ref}

\begin{thebibliography}{10}
\providecommand{\url}[1]{#1}
\csname url@samestyle\endcsname
\providecommand{\newblock}{\relax}
\providecommand{\bibinfo}[2]{#2}
\providecommand{\BIBentrySTDinterwordspacing}{\spaceskip=0pt\relax}
\providecommand{\BIBentryALTinterwordstretchfactor}{4}
\providecommand{\BIBentryALTinterwordspacing}{\spaceskip=\fontdimen2\font plus
\BIBentryALTinterwordstretchfactor\fontdimen3\font minus
  \fontdimen4\font\relax}
\providecommand{\BIBforeignlanguage}[2]{{%
\expandafter\ifx\csname l@#1\endcsname\relax
\typeout{** WARNING: IEEEtran.bst: No hyphenation pattern has been}%
\typeout{** loaded for the language `#1'. Using the pattern for}%
\typeout{** the default language instead.}%
\else
\language=\csname l@#1\endcsname
\fi
#2}}
\providecommand{\BIBdecl}{\relax}
\BIBdecl

\bibitem{Fan99}
P.~Z. Fan, N.~Suehiro, N.~Kuroyanagi, and X.~M. Deng, ``Class of binary
  sequences with zero correlation zone,'' \emph{Electron.\ Lett.}, vol.~35,
  no.~10, pp. 777--779, May 1999.

\bibitem{Tang10}
Y.-S. Tang, C.-Y. Chen, and C.-C. Chao, ``A novel construction of zero
  correlation zone sequences based on boolean functions,'' in \emph{Proc.\
  {IEEE} Int.\ Symp.\ on Spread Spectrum Tech.\ and Applicat.}, Taichung,
  Taiwan, Oct. 2010, pp. 198--203.

\bibitem{Torii04}
H.~Torii, M.~Nakamura, and N.~Suehiro, ``A new class of zero-correlation zone
  sequences,'' \emph{{IEEE} Trans. Inf. Theory}, vol.~50, pp. 559--565, Mar.
  2004.

\bibitem{Zhou08}
Z.~Zhou, X.~Tang, and G.~Gong, ``A new class of sequences with zero or low
  correlation zone based on interleaving technique,'' \emph{{IEEE} Trans. Inf.
  Theory}, vol.~54, pp. 4267--4273, Sep. 2008.

\bibitem{Liu14}
Z.~Liu, Y.~L. Guan, and U.~Parampalli, ``A new construction of zero correlation
  zone sequences from generalized {Reed-Muller} codes,'' in \emph{Proc.\ {IEEE}
  Inf. Theory Workshop}, Hobart, Australia, Nov. 2014, pp. 591--595.

\bibitem{Super_182}
C.-Y. Chen and S.-W. Wu, ``Golay complementary sequence sets with large zero
  correlation zones,'' \emph{{IEEE} Trans. Commun.}, vol.~66, no.~11, pp.
  5197--5204, Nov. 2018.

\bibitem{Pai_23}
C.-Y. Pai, Y.-J. Lin, and C.-Y. Chen, ``Optimal and almost-optimal
  {G}olay-{ZCZ} sequence sets with bounded {PAPRs},'' \emph{{IEEE} Trans.
  Commun.}, vol.~71, no.~2, pp. 728--740, Feb. 2023.

\bibitem{Yuan_05}
W.~Yuan, P.~Wang, and P.~Fan, ``Performance of multi-path {MIMO} channel
  estimation based on {ZCZ} training sequences,'' in \emph{Proc.\ Int.\ Symp.\
  on Microwave, Antenna, Propagation, and {EMC} Techno.\ for Wireless Commun.},
  Beijing, China, Sep. 2005, pp. 1542--1545.

\bibitem{Yang_02}
S.-A. Yang and J.~Wu, ``Optimal binary training sequence design for
  multiple-antenna systems over dispersive fading channels,'' \emph{{IEEE}
  Trans. Veh. Technol.}, vol.~51, no.~5, pp. 1271--1276, Sep. 2002.

\bibitem{Hu_17}
S.~Hu, Z.~Liu, Y.~L. Guan, C.~Jin, Y.~Huang, and J.-M. Wu, ``Training sequence
  design for efficient channel estimation in {MIMO-FBMC} systems,'' \emph{IEEE
  Access}, vol.~5, pp. 4747--4758, 2017.

\bibitem{Long98}
B.~Long, P.~Zhang, and J.~Hu, ``A generalized {QS-CDMA} system and the design
  of new spreading codes,'' \emph{{IEEE} Trans. Veh. Technol.}, vol.~47, pp.
  1268--1275, Nov. 1998.

\bibitem{Zhang_10}
W.~Zhang, F.~Zeng, X.~Long, and M.~Xie, ``Improved mutually orthogonal {ZCZ}
  polyphase sequence sets and their applications in {OFDM} frequency
  synchronization,'' in \emph{Proc.\ Int.\ Conf. on Wireless Commun. Netw. and
  Mobile Comput.}, Chengdu, China, Sep. 2010, pp. 1--5.

\bibitem{Zhang_12}
R.~Zhang, X.~Cheng, M.~Ma, and B.~Jiao, ``Interference-avoidance pilot design
  using {ZCZ} sequences for multi-cell {MIMO-OFDM} systems,'' in \emph{Proc.\
  IEEE Global Commun. Conf.}, Anaheim, CA, Dec. 2012, pp. 5056--5061.

\bibitem{SM_2}
R.~Y. Mesleh, H.~Haas, S.~Sinanovic, C.~W. Ahn, and S.~Yun, ``Spatial
  modulation,'' \emph{{IEEE} Trans. Veh. Technol.}, vol.~57, no.~4, pp.
  2228--2241, Jul. 2008.

\bibitem{SM_3}
M.~{Di Renzo}, H.~Haas, and P.~Grant, ``Spatial modulation for multiple-antenna
  wireless systems: {A} survey,'' \emph{{IEEE} Commun. Mag.}, vol.~49, no.~12,
  pp. 182--191, Dec. 2011.

\bibitem{SM_4}
P.~Yang, M.~{Di Renzo}, Y.~Xiao, S.~Li, and L.~Hanzo, ``Design guidelines for
  spatial modulation,'' \emph{IEEE Commun. Surv. Tut.}, vol.~17, no.~1, pp.
  6--26, May 1st Quart., 2015.

\bibitem{SM_6}
P.~Yang, Y.~Xiao, Y.~L. Guan, K.~Hari, A.~Chockalingam, S.~Sugiura, H.~Haas,
  M.~D. Renzo, C.~Masouros, Z.~Liu, L.~Xiao, S.~Li, and L.~Hanzo,
  ``{S}ingle-carrier {SM-MIMO}: {A} promising design for broadband large-scale
  antenna systems,'' \emph{IEEE Commun. Surv. Tut.}, vol.~18, no.~3, pp.
  1687--1716, 3rd Quart., 2016.

\bibitem{SM_5}
M.~Wen, B.~Zheng, K.~J. Kim, M.~{Di Renzo}, T.~A. Tsiftsis, K.~Chen, and
  N.~Al-Dhahir, ``A survey on spatial modulation in emerging wireless systems:
  Research progresses and applications,'' \emph{IEEE J. Sel. Areas Commun.},
  vol.~37, no.~9, pp. 1949--1972, Sep. 2019.

\bibitem{SM_Flat_2}
M.~{Di Renzo}, D.~D. Leonardis, F.~Graziosi, and H.~Haas, ``Space shift keying
  ({SSK}-) {MIMO} with practical channel estimates,'' \emph{{IEEE} Trans.
  Commun.}, vol.~60, no.~4, pp. 998--1012, Apr. 2012.

\bibitem{SM_Flat_3}
S.~Sugiura and L.~Hanzo, ``Effects of channel estimation on spatial
  modulation,'' \emph{{IEEE} Signal Process. Lett.}, vol.~19, no.~12, pp.
  805--808, Dec. 2012.

\bibitem{SM_timevaring_4}
X.~Wu, H.~Claussen, M.~{Di Renzo}, and H.~Haas, ``Channel estimation for
  spatial modulation,'' \emph{{IEEE} Trans. Commun.}, vol.~62, no.~12, pp.
  4362--4372, Dec. 2014.

\bibitem{SM_timevaring_5}
Y.~Akiba, T.~Ishihara, and S.~Sugiura, ``{V}ariable-block-length joint channel
  estimation and data detection for spatial modulation over time-varying
  channels,'' \emph{{IEEE} Trans. Veh. Technol.}, vol.~69, no.~11, pp.
  13\,964--13\,969, Nov. 2020.

\bibitem{CZCP-1st}
Z.~Liu, P.~Yang, Y.~L. Guan, and P.~Xiao, ``Cross {Z}-complementary pairs for
  optimal training in spatial modulation over frequency selective channels,''
  \emph{IEEE Trans. Signal Process.}, vol.~68, pp. 1529--1543, Feb. 2020.

\bibitem{Fan_20}
C.~Fan, D.~Zhang, and A.~R. Adhikary, ``New sets of binary cross
  {Z}-complementary sequence pairs,'' \emph{{IEEE} Commun. Lett.}, vol.~24,
  no.~8, pp. 1616--1620, Aug. 2020.

\bibitem{Adhikary_20}
A.~R. Adhikary, Z.~Zhou, Y.~Yang, and P.~Fan, ``Constructions of cross
  {Z}-complementary pairs with new lengths,'' \emph{IEEE Trans. Signal
  Process.}, vol.~68, pp. 4700--4712, 2020.

\bibitem{Huang_20}
Z.-M. Huang, C.-Y. Pai, and C.-Y. Chen, ``Binary cross {Z}-complementary pairs
  with flexible lengths from {B}oolean functions,'' \emph{{IEEE} Commun.
  Lett.}, vol.~25, no.~4, pp. 1057--1061, Apr. 2021.

\bibitem{Yang_21}
M.~Yang, S.~Tian, N.~Li, and A.~R. Adhikary, ``New sets of quadriphase cross
  {Z}-complementary pairs for preamble design in spatial modulation,''
  \emph{{IEEE} Signal Process. Lett.}, vol.~28, pp. 1240--1244, May 2021.

\bibitem{Zeng_22}
F.~Zeng, X.~He, Z.~Zhang, and L.~Yan, ``Quadriphase cross {Z}-complementary
  pairs for pilot sequence design in spatial modulation systems,'' \emph{{IEEE}
  Signal Process. Lett.}, vol.~29, pp. 508--512, Jan. 2022.

\bibitem{Das_22_ISIT}
S.~Das, A.~Banerjee, and Z.~Liu, ``New family of cross {Z}-complementary
  sequences with large {ZCZ} width,'' in \emph{Proc.\ {IEEE} Int.\ Symp.\ Inf.\
  Theory}, Espoo, Finland, Jun. 2022, pp. 522--527.

\bibitem{Zhang_22}
H.~Zhang, C.~Fan, Y.~Yang, and S.~Mesnager, ``New binary cross
  {Z}-complementary pairs with large {CZC} ratio,'' \emph{{IEEE} Trans. Inf.
  Theory}, vol.~69, no.~2, pp. 1328--1336, Feb. 2023.

\bibitem{Huang_22}
Z.-M. Huang, C.-Y. Pai, and C.-Y. Chen, ``Cross {Z}-complementary sets for
  training design in spatial modulation,'' \emph{{IEEE} Trans. Commun.},
  vol.~70, no.~8, pp. 5030--5045, Aug. 2022.

\bibitem{Huang_22_ISIT}
------, ``A novel construction of optimal cross {Z}-complementary sets based on
  generalized {B}oolean functions,'' in \emph{Proc.\ {IEEE} Int.\ Symp.\ Inf.\
  Theory}, Espoo, Finland, Jun. 2022, pp. 1725--1730.

\bibitem{Chang_67}
J.~A. Chang, ``Ternary sequence with zero correlation,'' \emph{Proc.\ {IEEE}},
  vol.~55, no.~7, pp. 1211--1213, Jul. 1967.

\bibitem{Xu_03}
S.~Xu and D.~Li, ``Ternary complementary orthogonal sequences with zero
  correlation window,'' in \emph{Proc.\ IEEE Int. Symp. Personal, Indoor and
  Mobile Radio Commun. (PIMRC)}, vol.~2, Beijing, China, Jan. 2003, pp.
  1669--1672.

\bibitem{Wu_06}
D.~Wu and P.~Spasojevic, ``Adaptive rate {QS}-{CDMA} {UWB} systems using
  ternary {OVSF} codes with a zero-correlation zone,'' in \emph{Proc.\ IEEE
  Wireless Commun. and Netw. Conf.}, vol.~2, Nevada, Apr. 2006, pp. 1068--1073.

\bibitem{Pai_22}
C.-Y. Pai and C.-Y. Chen, ``Two-dimensional {G}olay complementary array
  pairs/sets with bounded row and column sequence {PAPR}s,'' \emph{{IEEE}
  Trans. Commun.}, vol.~70, no.~6, pp. 3695--3707, Jun. 2022.

\bibitem{SM_ZPSC}
R.~{Rakshith}, K.~{V.S. Hari}, and L.~{Hanzo}, ``Spatial modulation aided
  zero-padded single carrier transmission for dispersive channels,''
  \emph{{IEEE} Trans. Commun.}, vol.~61, no.~6, pp. 2318--2329, Jun. 2013.

\end{thebibliography}
\IEEEtriggeratref{3}

\end{document}